\documentclass[twoside]{dis07}
\usepackage[latin1]{inputenc}
\usepackage[dvips]{graphicx,epsfig,color}
\usepackage{wrapfig,rotating}
\usepackage{cite,amssymb,amsmath,array}

\begin{document}
\title{{\normalsize\sl DESY 07/084     \hfill {\tt arXiv:yymm.nnnn}
\\ SFB/CPP-07-29 \hfill {   } }\\
Target Mass Corrections in Diffractive Scattering}

\author{J. Bl\"umlein$^1$, B. Geyer$^2$ and D. Robaschik$^3$
%
\thanks{This paper was supported in part by SFB-TR-9: Computergest\"utze 
Theoretische Teilchenphysik.}
%
\vspace{.3cm}\\
%
1- Deutsches Elektronen-Synchrotron, DESY \\
Platanenallee 6, D-15738 Zeuthen, Germany
%
\vspace{.1cm}\\
2- Center for Theoretical Studies and
Institute of Theoretical Physics Leipzig University,\\ Augustusplatz~10, 
D-04109~Leipzig,
Germany
\vspace{.1cm}\\
3- Brandenburgische Technische Universit\"at Cottbus,
Fakult\"at 1,\\ PF 101344, D--03013 Cottbus, Germany 
}

\maketitle

\begin{abstract}
We describe the twist-2 contributions to inclusive unpolarized and 
polarized deep-inelastic diffractive scattering in an operator approach.
The representation refers to the observed large rapidity gap but does
not require reference to a pomeron picture. We discuss both the case of
vanishing target mass $M$ and momentum transfer $t$ as well as the effects
at finite $t$ and $M$, which lead to modifications at large $\beta$ and 
low values of $Q^2$.
\end{abstract}

\section{Introduction}
Deep-inelastic diffractive scattering is one of the important scattering
processes in high-energy $ep$ scattering at HERA. In the small-$x$ domain 
$\sim 1/8$ of the events are due to this process. It is characterized by
inclusive hadron-production with a large rapidity gap between the outgoing
proton and all the remainder hadrons. In Refs.~\cite{BR1} two of the present
authors developed a description for this process based on the Compton operator
and using techniques known in non-forward scattering, cf. \cite{NF}, for the case
that the momentum transfer $t$ between the incoming and the outgoing proton and target
masses can be neglected. In the region of smaller values of $Q^2$ and large values 
of $\beta = x/x_P$ one expects both finite $t$ and $M^2$ effects which were worked out in
\cite{DTM} based on related investigations for the non-forward case \cite{GRE}, 
see also \cite{BM}. In this paper we summarize the main findings of these analyzes, cf. also 
\cite{url}.
\section{General Structure}
The hadronic tensor of the process is determined by three vectors $p_1, p_2, q$, the incoming and
outgoing proton momentum and the momentum transfer in the unpolarized 
case supplemented by the spin vector of the
initial proton $S$ in the polarized case. The following invariants are formed
\begin{eqnarray}
Q^2 =-q^2,~~ W:=(p_1+q)^2,~~x:=\frac{Q^2}{Q^+W^2-M^2},~~t:=(p_2-p_1)^2,~~
x_P := -\frac{2\eta}{2-\eta}  \geq x.  \nonumber
\end{eqnarray}
The hadronic tensors in case of the unpolarized and polarized cases are of the following form,
\cite{BR1}
{
\begin{eqnarray}
\label{eqD2}
W_{\mu\nu}^{{unp}} &=& \left( -g_{\mu\nu} + 
\frac{q_{\mu} q_{\nu}}{q^2}\right) 
{W_1}
           + \hat{p}_{1\mu}\hat{p}_{1\nu}
             \frac{{W_3}}{M^2}
           + \hat{p}_{2\mu}\hat{p}_{2\nu}
             \frac{{W_4}}{M^2}
           + \left(\hat{p}_{1\mu}\hat{p}_{2\nu}
           + \hat{p}_{2\mu}\hat{p}_{1\nu} \right)
             \frac{{W_5}}{M^2}
\nonumber
\end{eqnarray}
\hspace*{-1.5cm}
\begin{alignat}{10}
\label{eqH2}
W_{\mu\nu}^{{pol}}
 &=&~~
i \left[ \hat{p}_{1\mu} \hat{p}_{2\nu}-
\hat{p}_{1\nu} \hat{p}_{2\mu} \right] \varepsilon_{p_1,p_2,q,S}
&\frac{{\hat{W}_1}}{M^6}&~~
                &+&~~
i \left[ \hat{p}_{1\mu} \varepsilon_{\nu S p_1 q}
     - \hat{p}_{1\nu} \varepsilon_{\mu S p_1 q} \right]
&\frac{{\hat{W}_2}}{M^4}&
\nonumber \\ &+&~~
i \left[ \hat{p}_{2\mu} \varepsilon_{\nu S p_1 q}
     - \hat{p}_{2\nu} \varepsilon_{\mu S p_1 q} \right]
&\frac{{\hat{W}_3}}{M^4}&~~
                &+&~~
i \left[ \hat{p}_{1\mu} \varepsilon_{\nu S p_2 q}
     - \hat{p}_{1\nu} \varepsilon_{\mu S p_2 q} \right]
&\frac{{\hat{W}_4}}{M^4}&
\nonumber \\ &+&~~
i \left[ \hat{p}_{2\mu} \varepsilon_{\nu S p_2 q}
     - \hat{p}_{2\nu} \varepsilon_{\mu S p_2 q} \right]
&\frac{{\hat{W}_5}}{M^4}&~~
                &+&~~
i \left[ \hat{p}_{1\mu} \hat{\varepsilon}_{\nu p_1 p_2 S}
     - \hat{p}_{1\nu} \hat{\varepsilon}_{\mu p_1 p_2 S} \right]
&\frac{{\hat{W}_6}}{M^4}&
\nonumber\\ &+&~~
i \left[ \hat{p}_{2\mu} \hat{\varepsilon}_{\nu p_1 p_2 S}
     - \hat{p}_{2\nu} \hat{\varepsilon}_{\mu p_1 p_2 S} \right]
&\frac{{\hat{W}_7}}{M^4}&~~
                &+&~~  i \varepsilon_{\mu \nu q S}
&\frac{{\hat{W}_8}}{M^2}&~.
\nonumber
\end{alignat}
}

\vspace{0.1mm}\noindent
with $\hat{p}_{2\mu}, \hat{\varepsilon}_{\nu p_1 p_2 S}$, etc. the corresponding gauge-invariant 
completions. In general there are 4 unpolarized and 8 polarized structure functions in case of 
pure photon exchange.

The twist--2 contributions can be described applying the factorization theorem. Moreover,
A. Mueller's generalized optical theorem allows to turn the isolated final state proton into
an initial state anti-proton, being separated from the proton by $t$. In this way the 
{\it diffractive state} is formed, from which the hadronic tensor is obtained taking the forward
expectation value of the Compton-tensor. Evaluating the process further using the above kinematic 
variables we are led to a description of the diffractive scattering cross section which does
not require any reference to a pomeron picture, but is solely based on the presence of a large
rapidity gap.
\section{The Case \boldmath{$t= M^2 =O$}}
In this approximation the number of structure functions reduces to two unpolarized and two polarized
ones, because of the collinearity of $p_1$ and $p_2$,~\cite{BR1}. Due to this the diffractive 
state simplifies and leads to a Lorentz structure with lower complexity. For pure photon exchange only 
the structure 
functions $F_{1,2}$ resp. $g_{1,2}$ contribute, which in the twist-2 approximation obey a 
modified Callan--Gross relation
\begin{equation}
F_2(\beta, \eta, Q^2) = 2x F_1(\beta, \eta, Q^2)
\nonumber
\end{equation}
and the Wandzura-Wilczek, respectively. As shown in \cite{BR1}, the evolution equations, changing 
$x \rightarrow \beta$ are the same as for inclusive deep--inelastic scattering. To derive the 
diffractive evolution equations one considers
the evolution equations for non--forward scattering Ref.~[2b]
\begin{equation}
\mu^2 \frac{d}{d \mu^2} O^A(\kappa_+ \tilde{x}, \kappa_- \tilde{x}; \mu^2)
= \int D \kappa' \gamma^{AB}(\kappa_+,\kappa_-,\kappa'_+,\kappa_-'; \mu^2)
O^B(\kappa_+' \tilde{x}, \kappa_- '\tilde{x}; \mu^2)
\nonumber
\end{equation}
which turn into
\begin{equation}
\mu^2 \frac{d}{d \mu^2} f^A(\vartheta,\eta; \mu^2)
= \int_\vartheta^{-{\rm sign}(\vartheta)/\eta}
\frac{d\vartheta'}{\vartheta'}
P^{AB}
\left(\frac{\vartheta}{\vartheta'},
\mu^2\right)
f_B(\vartheta',\eta;\mu^2)
\nonumber
\end{equation}
in the case $t, M^2 \rightarrow 0$. The value of $\vartheta$ 
is determined by the absorptive condition as $\vartheta = 2 \beta$.
\section{Target Mass Corrections}
At low values of $Q^2$ and large values of $\beta$ both target mass and finite $t$--effects 
become important. As shown in  Ref.~\cite{DTM}, following \cite{GRE}, these effects have to be 
dealt with together. The method is a generalization of the treatment of target mass effects in 
\cite{GP} to the non--forward case.
The now more complicated diffractive states $\langle p_1,-p_2,t|$ imply that the pre-parton 
densities emerging in this case depend on two light--cone variables $z_\pm$, non of which can be 
integrated out. For further treatment we define the variables 
\begin{equation}
\vartheta = z_- + \frac{z_+}{\eta}, \hspace{1cm} \zeta = \frac{z_-}{\vartheta}. \nonumber
\end{equation}
The presence of the variable $\zeta$  implies that the full Lorentz 
structure outlined above contributes, assuming azimuthal angular integrals are not carried out.
Four unpolarized and eight polarized structure functions contribute.
The partonic description being possible in the case $t,M^2 \rightarrow 0$ at the level of
observables does not hold anymore in this case, since $p_1$ and $p_2$ are no longer collinear.
Instead, one has to perform definite integrals (the $\zeta$-integrals in \cite{DTM}) over  
pre-partonic two-particle correlation functions, which cannot be determined by experiment 
directly. The absorptive condition in the present case is given by
\begin{equation}
\vartheta = - \frac{2 \beta}{\kappa}~~
\frac{1}{1+ \sqrt{1+4 \beta^2 {\cal P}^2(\eta, \zeta,t)/Q^2}}~.
\nonumber
\end{equation}
Here $|{\cal P}(\eta, \zeta)|$ takes the role of the nucleon mass in the case of forward 
scattering. It holds ${\cal P}^2 = t(1-\zeta/\eta) + (4M^2-t)\zeta^2 \geq 0$.
As an example, we present the $M^2$ and $t$ corrections for the un-integrated unpolarized 
structure functions $F_{1,2}^a$ \cite{DTM}~:
\begin{eqnarray}
\label{F1x}
F^a_1(\vartheta,\zeta)
  & \equiv &
  \Phi_a^{(0)}(\vartheta,\zeta)
  + \frac{\kappa {\cal P}^2}{[(q{\cal P})^2-q^2{\cal P}^2]^{1/2}}
  \Phi_a^{(1)}(\vartheta, \zeta)
  + \frac{\kappa^2 [{\cal P}^2]^2}{(q{\cal P})^2-q^2{\cal P}^2}
  \Phi_a^{(2)}(\vartheta, \zeta)
\nonumber\\
\label{F2x}
F^a_2(\vartheta,\zeta)
  & \equiv &
  \Phi_a^{(0)}(\vartheta,\zeta)
  + \frac{3\kappa {\cal P}^2}{[(q{\cal P})^2-q^2{\cal P}^2]^{1/2}}
  \Phi_a^{(1)}(\vartheta,\zeta)
  + \frac{3\kappa^2  [{\cal P}^2]^2}{(q{\cal P})^2-q^2{\cal P}^2}
  \Phi_a^{(2)}(\vartheta,\zeta)
\nonumber
\end{eqnarray}
Here the $\zeta-$dependent distribution functions $\Phi_a^{(k)}(\vartheta,\zeta)$ are iterated
integrals of the correlation function $\Phi_a^{(0)}(\vartheta,\zeta)=f_a(\vartheta,\zeta)$, cf. 
\cite{DTM}. $a$ denotes the respective kinematic invariant, implying kinematic dependences
in general. 

Although no partonic description is obtained  one still may study, whether twist--2 relations 
between structure functions exist. 
In case of the Callan--Gross relation this is not expected, since it is absent also 
for forward scattering~\cite{GP}. However, the Wandzura--Wilczek relation between the twist--2 
contributions of the polarized structure functions $g_1$ and $g_2$ holds also in the diffractive 
case for finite values $M^2,t$, as in many other cases
\cite{BK,BT,WW}. Here the $\zeta-$integral can be carried out.\footnote{It would be interesting to 
see, whether the generalization of integral relations derived for the forward polarized case  
for the twist--2 and twist--3 contributions \cite{BK,BT} can be generalized to diffractive 
scattering for electro-weak boson exchange.} This is not the case for other structure functions.
Below this integral, however, all the different structure functions can be represented
by a {\it single} $\zeta-$dependent two--particle distribution function in the unpolarized 
and polarized case, respectively. The different $\zeta-$dependence of the respective pre-factors
and the fact that the $\zeta-$integral is definite prevents to access the corresponding 
pre--parton distribution functions. 

At the twist--2 level diffractive parton distributions exist whenever the $M^2, t \rightarrow 0$
approximation holds. For large values of $\beta$ and small values of $Q^2$ this is not the case.
This is also the kinematic region in which one expects higher twist operators to contribute in 
the light cone expansion.\footnote{For a phenomenological higher-twist approach see \cite{KGB}.}

The twist--2 scaling violations of the diffractive structure functions in case of $M^2,t$ being finite 
are different from those in the limit $M^2,t \rightarrow 0$. Unlike the case there, the 
non--forward evolution equations do not simplify in the same way and the $\zeta-$dependence will
remain here too.  
\section{Conclusions}
Deep-inelastic diffractive scattering can be described taking the 
expectation value of the {Compton Operator} {between the 
diffractive states} {$\langle p_1,p_2;t|$} {obtained by 
applying A. Mueller's generalized optical theorem.}
{In the limit} {$M^2, t \rightarrow 0$},
      {two polarized and two unpolarized structure 
functions}
      {contribute to the scattering cross section}
      {at twist $\tau = 2$}.     
{They are related by a} {modified Callan-Gross 
      relation (in 
      lowest order)}, {resp. the} {Wandzura-Wilzcek 
      relation in all orders.}
{Target mass corrections accounting for all $M^2, t$-effects are required
  in the region of large values of} {$\beta$} {and low values of}
  {$Q^2$}.
{The set of genuine diffractive structure 
functions becomes larger due to these effects:} {four unpolarized structure functions} 
{and} {eight polarized structure functions (with one relation) contribute.}
{These structure functions can be decomposed into} 
{generally different} {diffractive parton densities} {due 
to the $\zeta$--integral.}
{In the case of $M^2, t \rightarrow 0$ the} {scaling violations} {of the twist 
$\tau = 2$ contribution to the diffractive structure 
functions  are described by the} {evolution equations for  
forward scattering} {replacing} {$x \rightarrow \beta$.}
{The present approach results into a thorough 
description demanding a} {rapidity gap} { without any 
need to 
invoke a} {``pomeron''.}
%
%
%

\begin{footnotesize}

\end{footnotesize}

\begin{thebibliography}{99}
%
\bibitem{BR1}
  J.~Bl\"umlein and D.~Robaschik,
  Phys.\ Lett.\  B {\bf 517} (2001) 222.
  Phys.\ Rev.\  D {\bf 65} (2002) 096002.
%
\bibitem{NF}
  D.~M\"uller, D.~Robaschik, B.~Geyer, F.~M.~Dittes and J.~Horejsi,
  Fortsch.\ Phys.\  {\bf 42} (1994) 101.
  J.~Bl\"umlein, B.~Geyer and D.~Robaschik,
  Nucl.\ Phys.\  B {\bf 560} (1999) 283
;\\
  A.~V.~Belitsky and A.~V.~Radyushkin,
  Phys.\ Rept.\  {\bf 418} (2005) 1.
%
\bibitem{DTM}
  J.~Bl\"umlein, B.~Geyer and D.~Robaschik,
  Nucl.\ Phys.\  B {\bf 755} (2006) 112.
%
\bibitem{GRE}
  B.~Geyer, D.~Robaschik and J.~Eilers,
  Nucl.\ Phys.\  B {\bf 704} (2005) 279.
%
\bibitem{BM}
  A.~V.~Belitsky and D.~M\"uller,
  Phys.\ Lett.\  B {\bf 507} (2001) 173.
%
\bibitem{url} Slides: 
\verb$http://indico.cern.ch/contributionDisplay.py?contribId=133&sessionId=4&confId=9499$
%
\bibitem{GP}
  H.~Georgi and H.~D.~Politzer,
  Phys.\ Rev.\  D {\bf 14} (1976) 1829.
%
\bibitem{BK}
  J.~Bl\"umlein and N.~Kochelev,
  Phys.\ Lett.\  B {\bf 381} (1996) 296;
  Nucl.\ Phys.\  B {\bf 498} (1997) 285.
%
\bibitem{BT}
  J.~Bl\"umlein and A.~Tkabladze,
  Nucl.\ Phys.\  B {\bf 553} (1999) 427.
%
\bibitem{WW}
  J.~D.~Jackson, G.~G.~Ross and R.~G.~Roberts,
  Phys.\ Lett.\  B {\bf 226} (1989) 159;\\
  A.~Piccione and G.~Ridolfi,
  Nucl.\ Phys.\  B {\bf 513} (1998) 301;\\
  J.~Bl\"umlein and D.~Robaschik,
  Nucl.\ Phys.\  B {\bf 581} (2000) 449;\\
  B.~Geyer and M.~Lazar,
  Phys.\ Rev.\  D {\bf 63} (2001) 094003;\\
  J.~Bl\"umlein, V.~Ravindran and W.~L.~van Neerven,
  Phys.\ Rev.\  D {\bf 68} (2003) 114004;\\
  B.~Geyer and D.~Robaschik,
  Phys.\ Rev.\  D {\bf 71} (2005) 054018.
%
\bibitem{KGB}
  K.~Golec-Biernat and A.~Luszczak,
  arXiv:0704.1608 [hep-ph].
\end{thebibliography}
\end{document}